\documentclass[twocolumn]{revtex4}

\usepackage{graphicx}
\usepackage{dcolumn}
\usepackage{amsmath}

\begin{document}

\title{Massless modes and abelian gauge fields in multi-band
 superconductors
}

\author{Takashi Yanagisawa and Izumi Hase} 

\affiliation{Electronics and Photonics Research 
Institute, National Institute of Advanced Industrial Science and Technology (AIST),
Tsukuba Central 2, 1-1-1 Umezono, Tsukuba 305-8568, Japan
}


\begin{abstract}
In $N$-band superconductors, the $U(1)^N$ phase invariance is spontaneously 
broken.
We propose a model for $N$-band superconductors
where the phase differences between gaps are represented by abelian gauge fields.
This model corresponds to
an $SU(N)$ gauge theory with the abelian projection.
We show that there are massless modes as well as massive modes 
when the number of gaps $N$ is greater than 3.
There are $N-3$ massless modes and two massive modes near a minimum of
the Josephson potential when $N$ bands
are equivalent and Josephson couplings are frustrated.
The global symmetry $U(1)^{N-1}$ is broken down by the Josephson term to 
$U(1)^{N-3}$.
A non-trivial configuration of the gauge field, that is, a monopole
singularity of the gauge field results in a fractional 
quantum-flux vortex.
The fractional quantum-flux vortex
corresponds to a monopole in superconductors.
\end{abstract}


\maketitle

\section{Introduction}

Multi-phase physics is a new physics of multi-gap superconductors.
The ground state of superconductors has a long-range order by breaking 
the rotational invariance.
It is well known that the gapless Goldstone mode exists when the
continuous symmetry is spontaneously broken. 
For the complex order parameter, written as $\Delta=|\Delta|e^{i\theta}$,
any choice of $\theta$ would have exactly the same energy that implies
the existence of a massless Nambu-Goldstone boson. 
This changes qualitatively when the Coulomb interaction between electrons
is included.  The Coulomb repulsive interaction turns the massless
mode into a gapped plasma mode\cite{and58}.
This would change qualitatively in multi-gap superconductors because
an additional phase invariance will bring about novel phenomena.

The study of multi-gap superconductors stemmed from works by Kondo\cite{kon63}
and Suhl et al.\cite{suh59}.
The phase-difference mode between two gaps is now becoming an interesting 
topic\cite{leg66,gei67,izu90,agt99,sha02,tan02,bab02,gur03,sta10,tan10a,tan10b,
yan12,tan11,ota11,lin12,nit12,pla12}. 

The existence of fractionally quantized flux vortices is significant
in multi-band superconductors.  The fluctuation of phase-difference mode leads to
half-quantum flux vortices in two-gap superconductors\cite{izu90,tan02,bab02}. 
A generalization to a three-gap superconductor results in very
attractive features, that is,  chiral states with time-reversal
symmetry breaking and the existence of fractionally quantized 
vortices\cite{sta10,tan10a,tan10b,yan12}.

In this paper we investigate the multi-phase physics in multi-gap
superconductors.  We show that the phase-difference modes are
represented by gauge fields and this corresponds to the abelian
projection of an $SU(N)$ gauge theory.
We show that, in the case with more than four gaps, there are massless
modes when there is a frustration between Josephson couplings.
For an $N$-band superconductor with fully frustrated Josephson couplings,
there are $N-3$ massless modes and 2 massive modes, or $N-2$ massless
modes and 1 massive mode.

There is an interesting analogy between particle physics
and multi-band superconductivity.
The Higgs particle corresponds to a Higgs mode in superconductors where
the Higgs mode represents the fluctuation mode of the magnitude of the order
parameter $\Delta$\cite{end12}.  Thus, the energy gap of the Higgs mode is 
proportional to
$2|\Delta|$ being the inverse of the coherence length.
The dynamics of the Higgs mode will also be an interesting subject.
In an $N$-band superconductor, $N-1$ phase differene modes can be regarded
as the gauge fields, and the mass of the gauge field is given by the inverse
of the penetration depth.
The mass of the Higgs particle will correspond to the inverse of the
coherence length, and the masses of gauge bosons W and Z
correspond to the inverse of the penetration depth.
If we use $m_W\sim 80.41$GeV/$c^2$, $m_Z\sim 91.19$GeV/$c^2$ and
$m_H\sim 126$GeV/$c^2$, the Ginzburg-Landau parameter $\kappa$ of the
universe is
roughly estimated as
$\kappa =\lambda/\xi\sim m_{H}/m_{W,Z}\sim 1.5$.
When there is a Josephson term, this analogy will be modified because
the phase difference is fixed near a minimum of the Josephson potential.
This will be discussed in the following.

\section{Gauge-field representation}

Let us consider the Ginzburg-Landau free energy density of a two-band
superconductor without the Josephson term in a magnetic field:
\begin{eqnarray}
f&=& (\alpha_1|\psi_1|^2+\alpha_2|\psi_2|^2)+\frac{1}{2}\Big(\beta_1|\psi_1|^4
+\beta_2|\psi_2|^4\Big)\nonumber\\
&+& \frac{\hbar^2}{2m_1}\Big|\left(\nabla-i\frac{e^*}{\hbar c}{\bf A}\right)
\psi_1\Big|^2
+ \frac{\hbar^2}{2m_2}\Big|\left(\nabla-i\frac{e^*}{\hbar c}{\bf A}\right)
\psi_2\Big|^2\nonumber\\
&+& \frac{1}{8\pi}(\nabla\times{\bf A})^2,
\label{GL1}
\end{eqnarray}
where $\psi_j$ $(j=1,2)$ are the order parameters and $e^*=2e$.
Note that this functional is not invariant under the transformation:
\begin{equation}
\psi_j\rightarrow \exp\left(i\frac{e^*}{\hbar c}\bar{\theta_j}\right)\psi_j,~~~
{\bf A}\rightarrow {\bf A}+\nabla\chi.
\end{equation}
The functional is not invariant for any choice of $\chi$.
Let us adopt that the phase of $\psi_j$ is $\theta_j$: 
$\psi_j=e^{i\theta_j}|\psi_j|$, and define $\Phi=\theta_1+\theta_2$ and
$\varphi=\theta_1-\theta_2$.
We obtain the covariant derivative as
\begin{eqnarray}
\left(\nabla-i\frac{e^*}{\hbar c}{\bf A}\right)\psi_1
&=& e^{i\theta_1}\Big[\nabla-i\frac{e^*}{\hbar c}\left({\bf A}-\frac{\hbar c}{2e^*}
\nabla\Phi\right)\nonumber\\
&-&i\frac{e^*}{\hbar c}{\bf B}\Big]|\psi_1|,
\end{eqnarray}
and that for $\psi_2$ where the field
${\bf B}$ is the derivative of the phase difference $\varphi$,
\begin{equation}
{\bf B}= -\frac{\hbar c}{2e^*}\nabla\varphi.
\end{equation} 
We write ${\bf A}-\hbar c/(2e^*)\nabla\Phi$ as ${\bf A}$, and
then the free energy is written as
\begin{eqnarray}
f&=& (\alpha_1|\rho_1|^2+\alpha_2|\rho_2|^2)+\frac{1}{2}\Big(\beta_1|\rho_1|^4
+\beta_2|\rho_2|^4\Big)\nonumber\\
&+& \frac{\hbar^2}{2m_1}\Big|\left(\nabla-i\frac{e^*}{\hbar c}{\bf A}
-i\frac{e^*}{\hbar c}{\bf B}\right)\rho_1\Big|^2\nonumber\\
&+& \frac{\hbar^2}{2m_2}\Big|\left(\nabla-i\frac{e^*}{\hbar c}{\bf A}
+i\frac{e^*}{\hbar c}{\bf B}\right)\rho_2\Big|^2
+\frac{1}{8\pi}(\nabla\times{\bf A})^2,\nonumber\\
\label{GL2}
\end{eqnarray}
where $\rho_j=|\psi_j|$ ($j=1,2$).
When $m_1=m_2=m$, the kinetic part becomes 
\begin{equation}
f_{kin}= \frac{\hbar^2}{2m}\Big|\left(\nabla-i\frac{e^*}{\hbar c}{\bf A}\sigma_0
-i\frac{e^*}{\hbar c}{\bf B}\sigma_3\right)\psi\Big|^2+ \frac{1}{8\pi}
(\nabla\times{\bf A})^2,
\end{equation}
where
\begin{eqnarray}
\psi= \left(
\begin{array}{c}
\rho_1 \\
\rho_2 \\
\end{array}
\right).
\end{eqnarray}
$\sigma_0$ is unit matrix and $\sigma_3$ is the Pauli matrix.
This is a part of $SU(2)\times U(1)$ gauge theory (Weinberg-Salam
model).
The gauge field ${\bf A}$ appears as ${\bf A}-{\bf B}$ and
${\bf A}+{\bf B}$, or (after the gauge transformation) ${\bf A}$
and ${\bf A}+2{\bf B}\equiv {\bf Z}$.
Thus, the phase-difference fluctuation mode ${\bf B}$ appears as a linear
combination with ${\bf A}$.  The masses of gauge bosons are given by 
gap amplitudes
\begin{equation}
m_A = \frac{\hbar}{c}\frac{1}{\lambda_2}\propto |\psi_2|,~~~~
m_Z = \frac{\hbar}{c}\frac{1}{\lambda_1}\propto |\psi_1|,
\end{equation}
where $\lambda_1$ and $\lambda_2$ are
\begin{equation}
\lambda_1=\sqrt{\frac{1}{4\pi}\left(\frac{c}{e^*}\right)^2
\frac{m_1}{\rho_1^2}},~~~~
\lambda_2=\sqrt{\frac{1}{4\pi}\left(\frac{c}{e^*}\right)^2
\frac{m_2}{\rho_2^2}}.
\end{equation}
The penetration depth $\lambda_L$ is
given by
\begin{equation}
\frac{1}{\lambda_L^2}= \frac{1}{\lambda_1^2}+\frac{1}{\lambda_2^2}.
\end{equation}
This is because the current is given as (from the variational
condition $\delta f/\delta {\bf A}=0$)
\begin{eqnarray}
{\bf j}&=& \frac{1}{2}\hbar e^*\left( \frac{\rho_1^2}{m_1}
+\frac{\rho_2^2}{m_2}\right)\nabla\Phi+\frac{1}{2}\hbar e^*
\left( \frac{\rho_2^2}{m_2}-\frac{\rho_1^2}{m_1}\right)\nabla\varphi \nonumber\\
&-& \frac{(e^*)^2}{c}\left( \frac{\rho_1^2}{m_1}+\frac{\rho_2^2}{m_2}
\right){\bf A}.
\end{eqnarray}
This leads to
\begin{equation}
{\rm rot}{\bf j}= -\frac{(e^*)^2}{c}\left( \frac{\rho_1^2}{m_1}
+\frac{\rho_2^2}{m_2}\right) {\rm rot}{\bf A}.
\end{equation}
Since ${\rm rot}(\nabla\varphi)=0$, the phase-difference mode gives no
contribution to the penetration depth.
In the case of $\psi_2=0$ and $\psi_1\neq 0$, one gauge field ${\bf A}$
remains massless.

In the three-band case, the covariant derivative reads
\begin{equation}
D_{\mu}= \partial_{\mu}-i\frac{e^*}{\hbar c}A_{\mu}
-i\frac{e^*}{\hbar c}\sqrt{\frac{3}{2}}B_{\mu}^8
\lambda_8-i\frac{e^*}{\hbar c}\sqrt{\frac{3}{2}}B_{\mu}^3\lambda_3,
\end{equation}
where $\lambda_8$ and $\lambda_3$ are diagonal Gell-Mann matrices.
There are 2 phase-difference modes in the three-band case.
We have assumed that masses of all the bands are the same and
defined three phase variables
\begin{eqnarray}
\Phi&=& \theta_1+\theta_2+\theta_3, \\
\phi_8 &=& \theta_1+\theta_2-2\theta_3, \\
\phi_3 &=& \theta_1-\theta_2.
\end{eqnarray}
The gauge fields ${\bf B}_8$ and
${\bf B}_3$ are defined as
\begin{equation}
{\bf B}_8=-\frac{1}{3\sqrt{2}}\nabla\phi_8,~~
{\bf B}_3=-\frac{1}{\sqrt{6}}\nabla\phi_3,
\end{equation}
We can define three fields by a unitary transformation:
\begin{eqnarray}
Z_{\mu}&=& \frac{1}{\sqrt{3}}A_{\mu}+\frac{1}{\sqrt{6}}B_{\mu}^8
+\frac{1}{\sqrt{2}}B_{\mu}^3 \\
W_{\mu}&=& \frac{1}{\sqrt{3}}A_{\mu}+\frac{1}{\sqrt{6}}B_{\mu}^8
-\frac{1}{\sqrt{2}}B_{\mu}^3 \\
V_{\mu}&=& \frac{1}{\sqrt{3}}A_{\mu}-\sqrt{\frac{2}{3}}B_{\mu}^8.
\end{eqnarray}
The gauge fields $W$, $Z$ and $V$ acquire masses being proportional to
$|\psi_|$, $|\psi_2|$ and $|\psi_3|$, respectively.
In the case of one vanishing order parameter, we have one massless
gauge field.

It is straightforward to generalize the free energy to an $N$-band
superconductor. 
In the $N$-band case, we have $N-1$ phase-difference modes.
$N-1$ equals the rank of $SU(N)$.  The rank is the number of elements
of Cartan subalgebra, namely commutative generators.  
Let $t_1,\cdots,t_{N-1}$ be elements of the
Cartan subalgebra of $SU(N)$.  Then, the covariant derivative is
\begin{equation}
D_{\mu}= \partial_{\mu}-i\frac{e^*}{\hbar c}A_{\mu}
-i\frac{e^*}{\hbar c}\sum_{j=1}^{N-1}B_{\mu}^jt_j,
\end{equation}
and the free energy density (without the Josephson terms) is given by
\begin{equation}
f= \sum_j\alpha_j|\psi_j|^2+\frac{1}{2}\sum_j\beta_j|\psi_j|^4
+\frac{\hbar^2}{2m}|D_{\mu}\psi|^2+\frac{1}{8\pi}
(\nabla\times{\bf A})^2.
\end{equation}
Here, we adopted that masses are the same and 
$\psi=(\rho_1,\cdots,\rho_N)^t$ is a scalar field of order 
parameters.

We have shown that the phase-difference mode is none other than the gauge 
field. 
The phase-difference modes are represented by the diagonal part of the 
gauge field.  Let us write the gauge field $B_{\mu}$ in the form
\begin{equation}
B_{\mu}= \sum_{j=1}^{N-1}B_{\mu}^jt_j+\sum_{a=1}^{N^2-N}C_{\mu}^aX_a.
\end{equation}
$B_{\mu}^j$ $(j=1,\cdots,N-1)$ denote the diagonal elements of the
vector field and $C_{\mu}^a$ are the off-diagonal elements of the
vector field.  The phase-difference modes correspond to the diagonal part,
and this is the abelian projection of $SU(N)$ by 'tHooft\cite{tho81}.
A singularity of the gauge field $B_{\mu}^j$ appears as a monopole.
This singularity leads to a half-quantum flux vortex in a two-band
superconductor.  Fractionally quantized vortices arise as a result
of singularities of the gauge field.

\begin{figure}
\begin{center}
\includegraphics[width=5cm,angle=90]{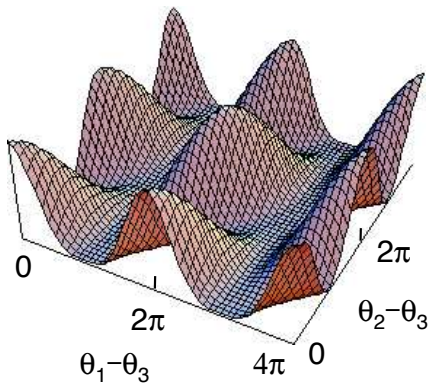}
\caption{
Josephson potential for the 4-band band as a function of
$\theta_1-\theta_3$ and $\theta_2-\theta_3$.
We set $\theta_1-\theta_3=\theta_2-\theta_4$ in the potential.
The flat minimum indicates an existence of zero mode. 
}
\label{4-band-v}
\end{center}
\end{figure}

\begin{figure}
\begin{center}
\includegraphics[width=2.5cm,angle=90]{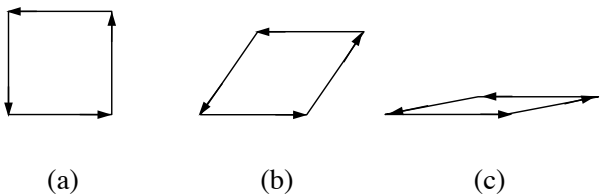}
\caption{
Spin configurations with $\sum_i{\bf S}_i=0$, forming squares, have the 
same energy for $N=4$.
The square of spins in (a) is mapped to (b) and (c) without finite excitation
energy.
This indicates that there is a massless mode.
The configuration in (a) corresponds to
$(\theta_1,\theta_2,\theta_3,\theta_4)=(0,\pi/2,\pi,3\pi/2)$.
}
\label{spin-config}
\end{center}
\end{figure}

\begin{figure}
\begin{center}
\includegraphics[width=7cm]{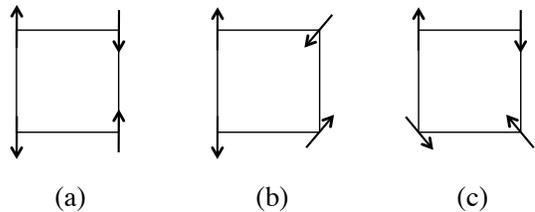}
\caption{
Configurations which have the same energy.
In (b) and (c) two spins can be rotated with the phase difference
fixed to be $\pi$.
}
\label{xy-spin}
\end{center}
\end{figure}

\section{Josephson term and massless modes}

There are $N-1$ gauge fields $B_{\mu}$ in the $N$-gap
superconductors.  We add the Josephson term to the free energy
functional, representing the pair transfer interactions between
different conduction bands\cite{kon63}.
The Josephson term is given as
\begin{equation}
f_J = -\sum_{i\neq j}\gamma_{ij}|\psi_i||\psi_j|\cos(\theta_i-\theta_j),
\end{equation}
where $\gamma_{ij}=\gamma_{ji}$ are chosen real.
This term obviously loses the gauge invariance of the free energy or
the Lagrangian because $\theta_i-\theta_j$
is not gauge invariant.
This indicates that the phase-difference modes acquire masses.
Hence, in the presence of the Josephson term, the phase-difference modes
are massive and there are excitation gaps.

To our surprise,
this would change qualitatively when $N$ is greater than 3.
We show that massless modes exist for an $N$-equivalent frustrated band
superconductor. 
Let us consider the Josephson potential given by
\begin{eqnarray}
V&=& \Gamma [ \cos(\theta_1-\theta_2)+\cos(\theta_1-\theta_3)
+\cos(\theta_1-\theta_4) \nonumber\\
&+& \cos(\theta_2-\theta_3)+\cos(\theta_2-\theta_4)
+\cos(\theta_3-\theta_4) ],
\end{eqnarray}
for $N=4$.  We assume that $\Gamma$ is positive: $\Gamma >0$ which indicates
that there is a frustration effect between Josephson couplings.
The ground states of this potential are degenerate.  For example,
the states with $(\theta_1,\theta_2,\theta_3,\theta_4)= (0,\pi/2,\pi,3\pi/2)$
and $(0,\pi,0,\pi)$ have the same energy. 
The Fig.1 shows $V$ as a function of $\theta_1-\theta_3$ and
$\theta_2-\theta_3$ in the case of $\theta_1-\theta_3=\theta_2-\theta_4$.
By expanding $V$ around a minimum $(0,\pi/2,\pi,3\pi/2)$,
 we find that there is one massless
mode and two massive modes.  
In fact, for $\theta_1-\theta_2=-\pi+\tilde{\eta}_1$, 
$\theta_2-\theta_4=-\pi+\tilde{\eta}_2$ and 
$\theta_2-\theta_3=-\pi/2+\tilde{\eta}_3$,
the potential $V$ is written as
\begin{equation}
V= \Gamma \Big[ -2+\frac{1}{2}\tilde{\eta}_1^2+\frac{1}{2}\tilde{\eta}_2^2+\cdots 
\Big],
\end{equation}
where the dots indicate higher order terms.
Missing of $\tilde{\eta}_3^2$ means that there is a massless mode 
and there remains a global $U(1)$ rotational symmetry, indicating that
 the ground states are continuously degenerate.
The gauge field corresponding to $\theta_2-\theta_3$ represents a massless
mode near $(\theta_1,\theta_2,\theta_3,\theta_4)=(0,\pi/2,\pi,3\pi/2)$.
One gauge symmetry is not broken and two gauge symmetries are broken for $N=4$.
The massive modes are represented by
linear combinations of $\theta_1-\theta_3$ and $\theta_2-\theta_4$.

We put
\begin{eqnarray}
\varphi_1 &=& \theta_1-\theta_2-\theta_3+\theta_4,\\ 
\varphi_2 &=& \theta_1+\theta_2-\theta_3-\theta_4,\\ 
\varphi_3 &=& \theta_1-\theta_2+\theta_3-\theta_4, 
\end{eqnarray}
then the covariant derivative is written as
\begin{equation}
D_{\mu}=\partial_{\mu}-i\frac{e^*}{\hbar c}A_{\mu}
-i\frac{e^*}{\hbar c}\sum_{j=1}^3t_jB_{\mu}^j,
\end{equation}
where
$A_{\mu}=(1/4)\partial_{\mu}(\theta_1+\theta_2+\theta_3+\theta_4)$, and
\begin{equation}
B_{\mu}^j = \frac{1}{4}\partial_{\mu}\varphi_j.
\end{equation}
The diagonal matrices of generators of $SU(4)$ are chosen as
\begin{eqnarray}
t_1&=&\left(
\begin{array}{cccc}
1 &  &  &  \\
  & -1 &   &  \\
  &    & -1 &  \\
  &    &    & 1 \\
\end{array}
\right),
t_2= \left(
\begin{array}{cccc}
1 &   &    &  \\
  & 1 &    &  \\
  &   & -1 &  \\
  &   &    & -1 \\
\end{array}
\right),\nonumber\\
t_3&=& \left(
\begin{array}{cccc}
1 &    &   &  \\
  & -1 &   &  \\
  &    & 1 &  \\
  &    &   & -1 \\
\end{array}
\right),
\end{eqnarray}
where $t_{\alpha}$ are normalized in the way 
${\rm Tr}t_{\alpha}t_{\beta}=4\delta_{\alpha\beta}$.
The potential $V$, near the minimum 
$(\varphi_1,\varphi_2,\varphi_3)=(0,-2\pi,\pi)$, is
\begin{equation}
V= \Gamma \Big[ -2+\frac{1}{4}(\eta_1^2+\eta_2^2)+\cdots \Big],
\end{equation}
for $\varphi_1=\eta_1$, $\varphi_2=-2\pi+\eta_2$ and $\varphi_3=\pi+\eta_3$.
Hence, the field $\varphi_3$, that is, $B_{\mu}^3$ represents a massless field.

We can generalize this argument for general $N$.  We show that
for $N\ge 4$, there exist always the massless modes for the potential
\begin{eqnarray}
V&=& \Gamma [ \cos(\theta_1-\theta_2)+\cos(\theta_1-\theta_3)+\cdots
+\cos(\theta_1-\theta_N)\nonumber\\
&+&\cdots+\cos(\theta_{N-1}-\theta_N) ].
\label{v-n}
\end{eqnarray}
For $\Gamma>0$, there are two massive modes and $N-3$ massless modes,
near the minimum 
$(\theta_1,\theta_2,\theta_3,\theta_4,\cdots)=(0,2\pi/N,4\pi/N,6\pi/N,\cdots)$.
We can check this for $N=4, 5, \cdots$.
We define $\theta_1-\theta_2=\Delta\theta+\xi_1$, 
$\theta_2-\theta_3=\Delta\theta+\xi_2$, $\cdots$, for $\Delta\theta=-2\pi/N$,
with the constraint $\xi_1+\xi_2+\cdots+\xi_N=0$.
By expanding $V$ in terms of $\xi_j$,  
$V$ is given by the quadratic form
$V/\Gamma=-N/2+\xi^tC\xi$ where $\xi^t=(\xi_1,\xi_2,\cdots,\xi_N)$.
The matrix $C$ is given by a Toeplitz matrix\cite{boe98} such as
\begin{eqnarray}
C =\left(
\begin{array}{ccccc}
c_0     & c_1  & c_2  & \cdots & c_{N-1}  \\
c_{N-1} & c_0  & c_1  & \cdots & \vdots  \\
\vdots  &        & \ddots  &  &  \\
c_2     & \cdots    &   &  & c_1 \\
c_1     & \cdots    &   &  & c_0 \\
\end{array}
\right).
\end{eqnarray}
The eigenvectors of $C$ are written as
$u_k=(1/\sqrt{N})(1,\zeta^k,\zeta^{2k},\cdots,\zeta^{(N-1)k})$ for 
$k=0,1,\dots,N-1$ with the
$N$-th root $\zeta=\exp(2\pi i/N)$.
The corresponding eigenvalues are
$\lambda_k=c_0+c_1\zeta^k+c_2\zeta^{2k}+\cdots+c_{N-1}\zeta^{(N-1)k}$
for $k=0,1,\cdots,N-1$. 
There are $N-3$ zero eigenvalues for $N$.
The number of massless modes increases as $N$ is increased.

When we expand the potential $V$ near the minimum
$(\theta_1,\theta_2,\theta_3,\theta_4)=(0,\pi,0,\pi)$, we obtain
\begin{equation}
V= \Gamma \Big[ -2+\frac{1}{2}\eta_1^2+\cdots \Big].
\end{equation}
This indicates that there are two massless modes and one massive mode.
When $N\ge 4$ and $N$ is even, we have $N-2$ massless modes and one
massive mode.
We show the number of massless modes in Table I and Table II.
We summarize the results as follows.
\\
\\
Proposition\\
{\em For the potential $V$ in eq.(\ref{v-n}) assuming $\Gamma>0$, 
there are $N-3$ massless modes and
2 massive modes near the minimum 
$(\theta_1,\theta_2,\cdots)=(0,2\pi/N,4\pi/N,\cdots)$, and there are
$N-2$ massless modes and one massive mode near
$(\theta_1,\theta_2,\cdots)=(0,\pi,0,\pi,0,\pi,\cdots)$.}
\\

The existence of massless modes can be understood by an analogy to a
classical spin model.  Let ${\bf s}_i$ ($i=1,\cdots,N$) be two-component vectors
with unit length $|{\bf S}_i|=1$.  Then, the potential $V$ is written as
\begin{equation}
V = \frac{\Gamma}{2}\Big[ \left(\sum_{i=1}^N{\bf S}_i\right)^2-N\Big].
\end{equation}
$V$ has a minimum $V_{min}=-\Gamma N/2$ for $\sum_i{\bf S}_i=0$.
Configurations under this condition have the same energy and
can be continuously mapped to each other with no excess energy.
At th $(\theta_1,\theta_2,\cdots)=(0,2\pi/N,4\pi/N,\cdots)$ with 
$V=-\Gamma N/2$,
satisfying $\sum_i{\bf S}_i=0$, the vectors ${\bf S}_i$ form a polygon.
The polygon can be deformed with the same energy (Fig.2).
For $N=4$, there is one mode of such deformation, which indicates that there
is one massless mode.
When $N$ increases by one, the number of zero modes increases by one.
Hence, we have $N-3$ massless modes and two massive modes for $N$.

At $(\theta_1,\theta_2,\cdots)=(0,\pi,0,\pi,0,\cdots)$ for $N\ge 4$ even,
satisfying also $\sum_i{\bf S}_i=0$, the polygon is folded into lines.
There are two zero modes in this configuration for $N=4$; 
we have rotational symmetry as shown in Fig.3 where two spins can rotate
with keeping the phase difference $\pi$.
We have two massless modes for $N=4$ and the number of massless modes
increases by one as $N$ increases by one.  Hence, there are
$N-2$ massless modes and one massive mode for $N$.

Let us generalize the potential as
\begin{eqnarray}
V&=& \Gamma[ \cos(\theta_1-\theta_2)+a\cos(\theta_1-\theta_3)
+\cos(\theta_1-\theta_4)\nonumber\\
&+&  \cos(\theta_2-\theta_3)
+a\cos(\theta_2-\theta_4)+\cos(\theta_3-\theta_4) ],\nonumber\\
\end{eqnarray}
for $N=4$ where $a$ is a real parameter.
When $a\ge 1$, $(\theta_1,\theta_2\theta_3,\theta_4)=(0,\pi/2,\pi,3\pi/2)$
is the ground state configuration and $V$ is expanded as
\begin{equation}
V= \Gamma \Big[ -2a+\frac{a}{4}(\eta_1^2+\eta_2^2) \Big].
\end{equation}
Then, there remains one massless mode for $a\ge 1$.
When $a\le 1 $, we expand $V$ near
$(\theta_1,\theta_2,\theta_3,\theta_4)=(0,\pi,0,\pi)$ to obtain
\begin{equation}
V= \Gamma\Big[ -4+2a+\frac{1}{4}\left( (1-a)(\eta_1^2+\eta_2^2)+2\eta_3^2\right) 
\Big].
\end{equation}
In this case there is no massless mode and mass for $\eta_1$ and $\eta_2$
modes is proportional to $1-a$.  When $a$ is close to 1, we have an
excitation state with small excitation energy.

\begin{table}
\caption{The number of massive and massless modes for
$|\Delta\theta|=2\pi/N$ for the potential with equivalent interactions.
}
\begin{center}
\begin{tabular}{cccc}
\hline
$N$    & massive modes & massles modes  & total \\
\hline
2   &  1  &  0    &  1  \\
3   &  2  &  0    &  2 \\
4   &  2  &  1    &  3  \\
5   &  2  &  2    &  4  \\
6   &  2  &  3    &  5  \\
$N$ &  2  & $N$-3 & $N$-1 \\
\hline
\end{tabular}
\end{center}
\end{table}

\begin{table}
\caption{The number of massive and massless modes for
$|\Delta\theta=\pi$ for the potential with equivalent interactions,
where $N$ is even.
}
\begin{center}
\begin{tabular}{cccc}
\hline
$N$    & massive modes & massles modes  & total \\
\hline
4   &  1  &  2   &  3  \\
6   &  1  &  4   &  5  \\
$N$ (even)   &  1  & $N$-2  &  $N$-1 \\
\hline
\end{tabular}
\end{center}
\end{table}

\section{Non-trivial configuration of gauge field}

In the following let us discuss a role of the gauge field
${\bf B}$.
We show that a monopole singularity of the gauge field ${\bf B}$ leads to
a fractional-quantum flux vortex.
In a two-band superconductor with $|\psi_1|=|\psi_2|$ and $m_1=m_2$,
the phase difference $\phi$ satisfies the sine-Gordon equation:
$d^2\phi/dx^2=\kappa\sin\phi$,
where we assume that the phase difference $\phi$ has spatial
dependence only in one direction, for example, in x direction.
We use the boundary condition such that $\phi\rightarrow 0$ as
$x\rightarrow -\infty$ and $\phi\rightarrow 2\pi$ as $x\rightarrow\infty$.
Then we have a kink solution:
\begin{equation}
\phi= \pi+2\sin^{-1}(\tanh(\sqrt{\kappa}x)),
\end{equation}
where we can adopt that $\kappa$ is positive because the sign of $\kappa$
does not matter since we can change the sign of $\sin\phi$ by shifting
the variable $\phi$.
The phase difference $\phi$ changes from 0 to $2\pi$ across the
kink.  This means that $\theta_1$ changes from 0 to $\pi$ and at the
same time $\theta_2$ changes from 0 to $-\pi$.
In this case, a half-quantum-flux vortex exists at the end of the kink.
This is shown in Fig.4 where the half-quantum vortex is at the
edge of the cut (kink).
A net change of $\theta_1$ is $2\pi$ by a counterclockwise
encirclement of the vortex, and that of $\theta_2$ vanishes.  Then,
we have a half-quantum flux vortex.
The half-flux vortex has also been investigated in the study of p-wave
superconductivity\cite{vol03,kee00,jan11}.
In the case of chiral p-wave superconductors, however, the singularity of
U(1) phase is canceled by the kink structure of the d-vector.
This is the difference between the two-band superconductor and the
p-wave superconductor.

\begin{figure}
\begin{center}
\includegraphics[width=3cm,angle=90]{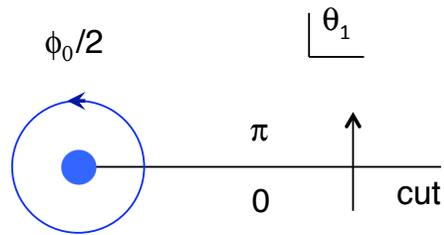}
\caption{
Half-quantum flux vortex with line singularity.
The phase variables $\theta_1$ with a line singularity.
}
\label{half-cut}
\end{center}
\end{figure}

The phase-difference gauge field ${\bf B}$ is defined as
\begin{equation}
{\bf B}= -\frac{1}{2}\nabla\bar{\phi} = -\frac{\hbar c}{2e^*}\nabla\phi.
\end{equation}
The half-quantum vortex can be interpreted as a monopole.
Let us assume that there is a cut, namely, kink on the real axis
for $x>0$.  The phase $\theta_1$ is represented by
\begin{equation}
\theta_1 = -\frac{1}{2}{\rm Im}\log\zeta+\pi,
\end{equation}
where
$\zeta= x+iy$.
The singularity of $\theta_j$ can be transferred to a singularity of the
gauge field by a gauge transformation.
We consider the case $\theta_2=-\theta_1$: $\phi=2\theta_1$.  Then we have
\begin{equation}
{\bf B}= -\frac{\hbar c}{2e^*}\nabla\phi= -\frac{\hbar c}{e^*}\frac{1}{2}
\left( \frac{y}{x^2+y^2},-\frac{x}{x^2+y^2},0 \right).
\end{equation}
Thus, when the gauge field ${\bf B}$ has a monopole-type singularity,
the vortex with half-quantum flux exists in two-gap superconductors.

Let us consider the fictitious z axis perpendicular to the x-y plane.
The gauge potential (1-form) is given by
\begin{equation}
\Omega_{\pm}= -\frac{1}{2}\frac{1}{r(z\pm r)}(ydx-xdy)=\frac{1}{2}(\pm 1
-\cos\theta)d\varphi,
\end{equation}
where $r=\sqrt{x^2+y^2+z^2}$, and $\theta$ and $\varphi$ are Euler angles.
$\Omega_{\pm}$ correspond to the gauge potential in the upper and lower
hemisphere $H_{\pm}$, respectively.  $\Omega_{\pm}$ are connected by
$\Omega_+ = \Omega_-+d\varphi$.
The components of $\Omega_+$ are
\begin{equation}
\Omega_{\mu}= \frac{1}{2}(1-\cos\theta)\partial_{\mu}\varphi.
\end{equation}
At $z=0$, $\Omega_{\mu}$ coincides with the gauge field for half-quantum
vortex.  If we identify $\varphi$ with $\phi$, we obtain
\begin{equation}
{\bf B}= \frac{\hbar c}{e^*}{\bf \Omega},
\end{equation}
at $\theta=\pi/2$.
$\{\Omega_{\pm}\}$ is the U(1) bundle $P$ over the sphere $S^2$. 
The Chern class is defined as
\begin{equation}
c_1(P)= -\frac{1}{2\pi}F=-\frac{1}{2\pi}d\Omega_+.
\end{equation}
The Chern number is given as
\begin{eqnarray}
C_1&=& \int_{S^2}c_1 = -\frac{1}{2\pi}\int_{S^2}F\nonumber\\
&=& -\frac{1}{2\pi}\left( \int_{H_+}d\Omega_++\int_{H_-}d\Omega_-\right)=1.
\end{eqnarray}
In general, the gauge field ${\bf B}$ has the integer Chern number:
$C_1=n$.
For $n$ odd, we have a half-quantum flux vortex.

\section{Application and discussion}

\begin{figure}
\begin{center}
\includegraphics[width=6cm]{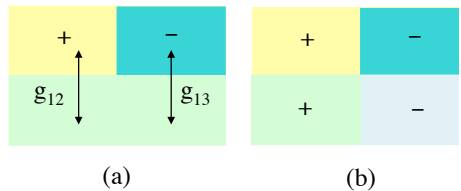}
\caption{
Schematic pictures of (a) $s_{\pm}-s$ junction and (b) $s_{\pm}-s_{\pm}$
junction.  In (a) $g_{12}$ is the Josephson coupling between the $s$-wave
superconductor and $s_+$-band of $s_{\pm}$ superconductor.  $g_{13}$ is
that for $s$-wave and $s_-$-band.
}
\label{s-s-junc}
\end{center}
\end{figure}

We point out that we can apply our theory to a junction of two
superconductors.  For example, a junction of two $s_{\pm}$ superconductors
may be described by an $N=4$ model in this paper where we set
$(\theta_1,\theta_2,\theta_3,\theta_4)=(0,\pi,0,\pi)$ (in Fig.3).
A zero energy mode may exist in this junction.
The Fig.\ref{s-s-junc} shows 
$s_{\pm}-s$ and $s_{\pm}-s_{\pm}$ junctions.
In a junction of $s_{\pm}$-wave and $s$-wave superconductors, the
potential energy is proportional to $|g_{12}-g_{13}|\varphi^2$ where
$g_{12}$ and $g_{13}$ are Josephson coupling between $s$- and 
$s_{\pm}$-wave superconductors and $\varphi$ is the relative phase.
When $g_{12}=g_{13}$, there is a massless mode in this system.

There is now a controversy about the symmetry of gap function in
iron-based superconductors, that is, $s_{++}$- or $s_{\pm}$-wave
symmetry\cite{shi09,yan09,yan10,kur08,maz08,kon10,sat09}.
We propose a method to determine the symmetry, $s_{++}$ or $s_{\pm}$,
by a junction which consists of two superconductors where both are 
iron-based superconductors or one is replaced by an $s$-wave superconductor.
We indicate a possibility that there is a low-energy excited state 
if the gap symmetry is $s_{\pm}$. 
In the case of Fig.5(a), the energy with respect to the phase difference
variable $\varphi$ is
\begin{equation}
E_{\varphi}= -2\gamma_{+-}|\Delta_+||\Delta_-|
+2(\gamma_+|\Delta_+|-\gamma_-|\Delta_-|)|\Delta|\cos(\varphi),
\end{equation}
where $\Delta_+$, $\Delta_-$ and $\Delta$ are gap functions of $s_+$-,
$s_-$-components of an $s_{\pm}$ superconductor and an $s$-wave
superconductor, respectively. 
$\varphi$ is the phase difference between $\Delta_+$ and $\Delta$. 
$\gamma_+$ and $\gamma_-$ are Josephson couplings
between $\Delta_+$ and $\Delta$, $\Delta_-$ and $\Delta$, respectively, and
$\gamma_{+-}$ is that between $\Delta_+$ and $\Delta_-$.
We adopt that $\gamma_+$ and $\gamma_-$ are positive.
Parameters $\gamma_+$ and $\gamma_-$ are dependent on a junction between
$s_{\pm}$- and $s$-wave superconductors and are probably controllable
artficially.
When $\gamma_+|\Delta_+|-\gamma_-|\Delta_-|$ is small or vanishing, we have a
low-energy excited state.  The existence of low-energy state will 
yield some structure in the density of states and will
be observed by several experiments such as the specific heat measurement
or the tunneling spectroscopy.
In contrast,
when the gap symmetry is $s_{++}$ in stead of $s_{\pm}$, the second
term in $E_{\varphi}$ is given by 
$2(\gamma_+|\Delta_+|+\gamma_-|\Delta_-|)|\Delta|\cos(\varphi)$.
In this case, we have no small excitation energy.
Hence, we can determine the symmetry $s_{\pm}$ or $s_{++}$ by the
existence of a low-energy excited state.
This argument can be also applied to the junction shown in Fig.5(b).

From our viewpoint, the phase-difference mode is represented by the vector
field which is regarded as the gauge field.
The fluctuation of the phase-difference mode is represented by the dynamics
of gauge field. The non-trivial topology of the gauge field corresponds to
the fractional-quantum flux vortex, and thus a multi-band superconductor
is regarded as a topological superconductor. 
We can generalize the free energy in eq.(\ref{GL2}) to 
\begin{eqnarray}
f&=& (\alpha_1|\psi_1|^2+\alpha_2|\psi_2|^2)+\frac{1}{2}\Big(\beta_1|\psi_1|^4
+\beta_2|\psi_2|^4 \Big)\nonumber\\
&+& \frac{\hbar^2}{2m_1}\Big|\left(\nabla-i\frac{e^*}{\hbar c}{\bf A}
-i\frac{e^*}{\hbar c}{\bf B}\right)\psi_1\Big|^2\nonumber\\
&+& \frac{\hbar^2}{2m_2}\Big|\left(\nabla-i\frac{e^*}{\hbar c}{\bf A}
+i\frac{e^*}{\hbar c}{\bf B}\right)\psi_2\Big|^2
+\frac{1}{8\pi}(\nabla\times{\bf A})^2,\nonumber\\
\label{GL3}
\end{eqnarray}
where $\psi_j$ are complex-valued fields.
This model has the gauge invariance.
It is not clear whether the field theory in eq.(\ref{GL2}) is
equivalent to that in eq.(\ref{GL3}).
It may be interesting to investigate the field theory in eq.(\ref{GL3})
and relevance to real superconductivity.

\section{Summary}


We have shown that
the phase difference modes are represented as gauge fields.
The action of the $N$-band superconductor is given by the abelian
projection of an $SU(N)$ gauge theory. 
In general,  the pair-transfer term (Josephson term)
breaks the gauge invariance and phase difference modes acquire masses 
because the phase differences are fixed
near minimums of the Josephson potential.
When $N$ is greater than 3, there are, however, massless modes when
$N$ bands are equivalent.
There are $N-3$ massless modes and 2 massive modes near the minimum
$(\theta_1,\theta_2,\cdots)=(0,2\pi/N,4\pi/N,\cdots)$, and
$N-2$ massless modes and 1 massive mode near 
$(\theta_1,\theta_2,\cdots)=(0,\pi,0,\pi,\cdots)$.

In iron-based superconductors the unconventional isotope effect appears
when the signs of two gap functions are opposite to each 
other\cite{shi09,yan09,yan10}.  
This is clearly a multi-band effect and suggests that the Cooper pairs
in  some of iron pnictides have $s_{\pm}$ symmetry.
A junction of two $s_{\pm}$ superconductors may be described by an
$N=4$ model proposed in this paper.
Therefore, there is a possibility that a new mode with zero or low 
excitation energy mode exits in the junction.  
 
A non-trivial configuration of the gauge fields of phase-difference modes
results in the existence of a fractional-quantum-flux vortex. 
The gauge field has a monopole singularity with the integer Chern
number.
In this sense, a multi-band superconductor can be a topological
superconductor.

We thank K. Yamaji, Y. Tanaka and K. Odagiri for stimulating
discussions.


\end{document}